\documentclass[twocolumn,twoside,slac_two]{revtex4}
\usepackage{graphicx}
\usepackage{fancyhdr}
\usepackage{subfig}
\begin{document}

\title{The spectral energy distribution of gamma-faint compact radio sources}

%

\author{C. S. Chang$^{1}$}

\author{E. Ros$^{1,2}$}

\author{M. Kadler$^{3,4,5}$}

\author{the \textit{Fermi} LAT Collaboration}

\author{the F-GAMMA team\footnote{The F-GAMMA team (in alphabetical order): E. Angelakis, L. Fuhrmann, T. P. Krichbaum, S. Larsson, N. Marchili, G. Nestoras, H. Ungerechts, J. A. Zensus et al.} }
\affiliation{$^{1}$Max-Planck-Institut f\"ur Radioastronomie, Auf dem H\"ugel 69, D-53121 Bonn, Germany}
\affiliation{$^{2}$Departament d'Astronomia i Astrof\'{\i}sica, Universitat de Val\`encia, E-46100 Burjassot, Spain }
\affiliation{$^{3}$Dr. Remeis-Sternwarte \& ECAP, Sternwartstr. 7, D-96049 Bamberg, Germany }
\affiliation{$^{4}$CRESST/NASA Goddard Space Flight Center, Greenbelt, MD 20771, USA}
\affiliation{$^{5}$USRA, 10211 Wincopin Circle, Suite 500 Columbia, MD 21044, USA}

\begin{abstract}
MOJAVE is a VLBI program which monitors a statistically complete, radio-selected sample of 135 relativistically beamed, flat-spectrum active galactic nuclei for over more than a decade. In order to understand the high-energy behavior of this radio complete sample, we are performing \textit{Swift} fill-in observations on the complete MOJAVE-I sample since 2007. The complete study of the spectral energy distribution from radio to X-ray bands on this radio-selected sample will provide us an opportunity to understand the nature of AGN. Here we present the preliminary results of the spectral energy distributions of six gamma-quiet or faint sources from this project: NRAO\,140, PKS\,0403$-$13, PKS\,B0422$+$004, PKS\,0823$+$033, 3C\,309.1, and 3C\,380.
\end{abstract}

\maketitle

\thispagestyle{fancy}

\section{Introduction}
The MOJAVE (monitoring of jets in active galactic nuclei with VLBA experiments) program \citep{lister09a} has been monitoring a statistically complete, radio flux limited sample \citep{lister05} of 135 radio loud active galactic nuclei (AGN) for more than a decade. This monitoring program provides the parsec-scale morphology of AGN jets evolving with time. After \textit{Fermi}'s launch in June 2008, we have the opportunity to investigate the relation between the parsec-scale radio-jet properties and the gamma-ray emission. First results of the correlation between gamma and radio properties have been presented in \citep{lister09c}-\citep{savolainen09}. The gamma-ray emission is thought to be generated close to the super massive black holes hosted in AGN. Moreover, with the simultaneous multiwavelength observations of AGN, we are able to investigate the relation between parsec-scale jets and broadband spectral energy distribution (SED) of AGN. 

\medskip

To understand the high-energy behavior of the MOJAVE sample, we are performing \textit{Swift} fill-in observations since 2007 on all the objects in the sample. Our aim is to study the SED of the radio-selected AGN sample from radio to gamma-ray, search for correlations, and to picture the physical mechanisms behind. In this paper, we selected six radio-bright AGN which are faint or quiet in the preliminary LAT first-year catalog, and which were not detected by EGRET. Studying those sources could help us to understand the connection between radio and gamma-ray emissions, and the reason which makes those bright radio sources to be faint in gamma-rays.  Here we present and discuss the preliminary results obtained by the broadband study of the six sources.

\bigskip
\bigskip

\section{Preliminary results}

\begin{table*}[Ht]
\caption{Source properties}
\begin{tabular*}{1\textwidth}{@{}c  l c  c c c  l@{}}
\hline
\textbf{IAU B\,1950 name} & \textbf{Alt. name} & \textbf{Source type} & \textbf{z} & $\beta_{\mathrm{app}}$$^\mathrm{a}$ & \textbf{LAT 1st-year$^\mathrm{b}$} & \textbf{pc-scale structure}$^\mathrm{c}$  \\
\hline
0333+321 & NRAO 140 & LPRQ$^\mathrm{d}$ & 1.258 & 12.9 $\pm$ 0.3 & Y & core-jet over 20\,mas (168\,pc) \\
0403$-$132 & PKS 0403$-$13 & HPRQ$^\mathrm{e}$ & 0.571 & 21.0 $\pm$ 1.0 & Y & core-jet over 15\,mas (97.7\,pc) \\
0422+004 & PKS B0422$+$004 & BL Lac & - & No redshift$^\mathrm{f}$ & Y & core-jet over 10\,mas \\
0823+033 & PKS 0823$+$033 & BL Lac & 0.506 & 15.0 $\pm$ 1.0 & Y & core-jet over 15\,mas (91.8\,pc) \\
1458+718 & 3C 309.1 & LPRQ$^\mathrm{d}$ & 0.905 & \,\,\,8.0 $\pm$ 3.0 & N & core-jet over 30\,mas (235\,pc) \\
1828+487 & 3C 380 & LPRQ$^\mathrm{d}$ & 0.692 & 13.1 $\pm$ 0.4 & Y & core-jet over 25\,mas (178\,pc) \\
\hline
\multicolumn{6}{@{}l@{}}{\footnotesize{$^\mathrm{a}$ Apparent projected speed in units of the speed of light (values from MOJAVE, see \citep{lister09b}). }} \\
\multicolumn{6}{@{}l@{}}{\footnotesize{$^\mathrm{b}$ Preliminary LAT first-year catalog. }} \\
\multicolumn{6}{@{}l@{}}{\footnotesize{$^\mathrm{c}$ pc-scale structure from \citep{lister05}. }} \\
\multicolumn{6}{@{}l@{}}{\footnotesize{$^\mathrm{d}$ Low polarization radio quasar. }} \\
\multicolumn{6}{@{}l@{}}{\footnotesize{$^\mathrm{e}$ High polarization radio quasar. }} \\
\multicolumn{6}{@{}l@{}}{\footnotesize{$^\mathrm{f}$ The redshift of this source is not available. The sky proper motion is 0.38 $\pm$ 0.08 mas\,yr$^{-1}$. }} \\
\end{tabular*}
\label{tab:src_property}
\end{table*}

We show the preliminary results of the following sources: NRAO\,140, PKS\,0403$-$13, PKS\,B0422$+$004, PKS\,0823$+$033, 3C\,309.1, and 3C\,380. All sources present a compact core-jet structure, and the source properties are shown in Table \ref{tab:src_property}. 

To construct the SED of the 6 sources we collected datasets obtained during \textit{Fermi}'s mission, which started on 11 June 2008. The simultaneous data we use in this paper include: the University of Michigan Radio Astronomy Observatory (UMRAO), VLBA from the MOJAVE program, Effelsberg 100-m telescope, and the IRAM Plateau de Bure Interferometer (all four in radio), \textit{Swift} UVOT/XRT (optical/UV and X-ray), and \textit{Fermi} LAT (gamma-ray) datasets. Effelsberg and PdB data are obtained in the framework of the F-GAMMA project \citep{fuhrmann07}. The \textit{Swift} data were analyzed by using NASA's HEASOFT software in its version 6.6.3. The optical/UV flux densities of \textit{Swift} UVOT are obtained using aperture photometry, and the X-ray flux densities of \textit{Swift} XRT data are estimated by using \texttt{XSPEC} version 12.5.0 to fit the spectra. \textit{Fermi} LAT data were analyzed by using \textit{Fermi} Science Tools version v9r10 and instrumental response functions (IRFs) version \texttt{P6\_V3\_DIFFUSE}. For the broadband SED, we included archival values from NASA extragalactic database (NED).

\begin{figure*}
 \centering
 \includegraphics[angle=270, width=0.75\textwidth]{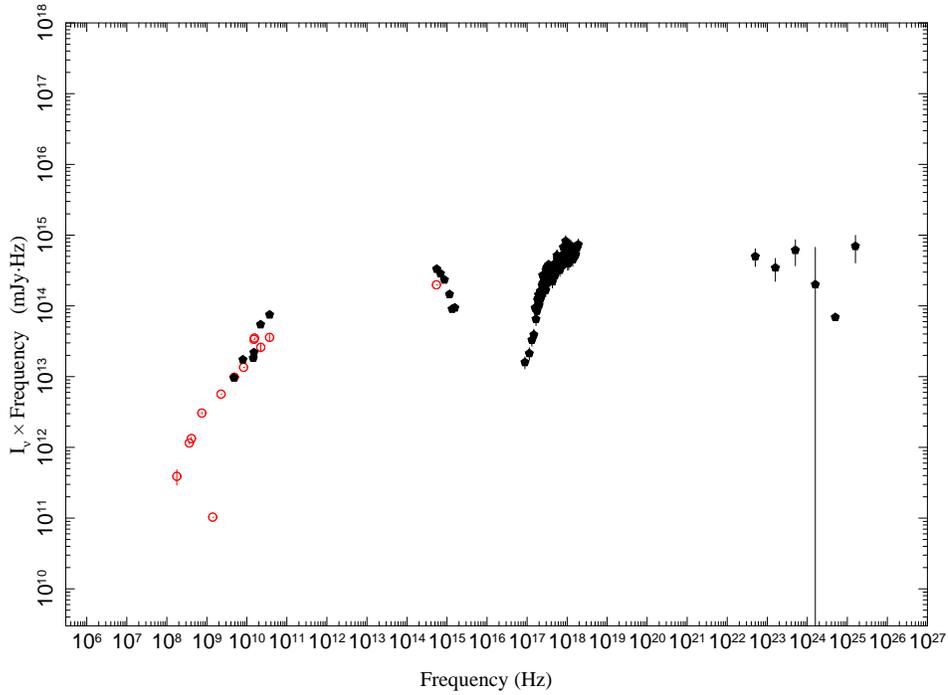}
 \caption{Broadband SED from radio to gamma-ray of NRAO 140. The data obtained before the \textit{Fermi} launch are red open circles (mostly taken from the NED), while the ones obtained later are black filled pentagons. The black data points are from: UMRAO, MOJAVE, Effelsberg, IRAM Plateau de Bure Interferometer, \textit{Swift} UVOT/XRT, and \textit{Fermi} LAT.  }
 \label{fig:sed2}
\end{figure*}

\begin{figure*}
 \centering
 \includegraphics[angle=270, width=0.75\textwidth]{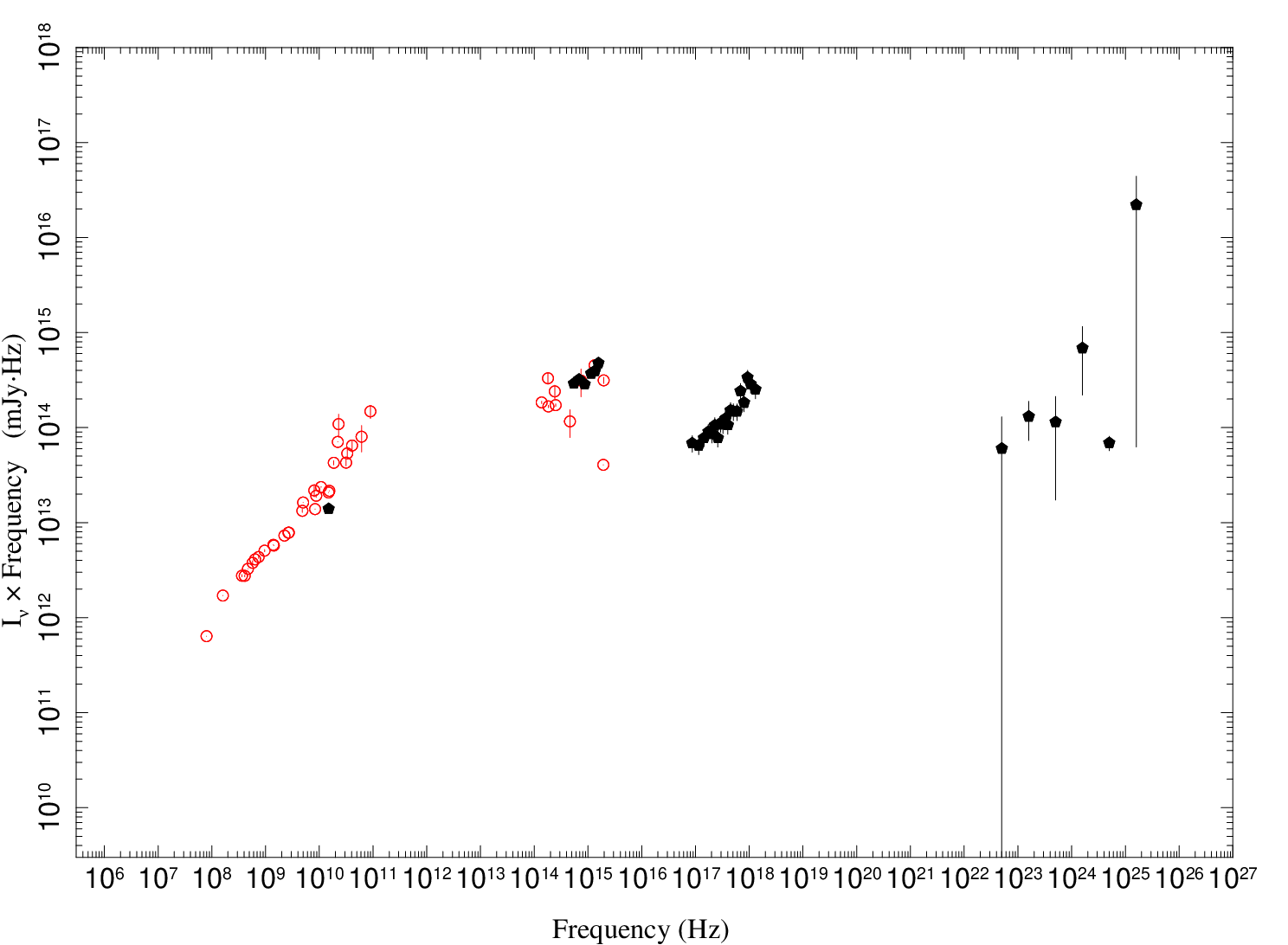}
 \caption{Broadband SED from radio to gamma-ray of PKS\,0403$-$13. The data obtained before the \textit{Fermi} launch are red open circles (mostly taken from the NED), while the ones obtained later are black filled pentagons. The black data points are from: UMRAO, MOJAVE, \textit{Swift} UVOT/XRT, and \textit{Fermi} LAT. }
 \label{fig:sed3}
\end{figure*}

\begin{figure*}
 \centering
 \includegraphics[angle=270, width=0.75\textwidth]{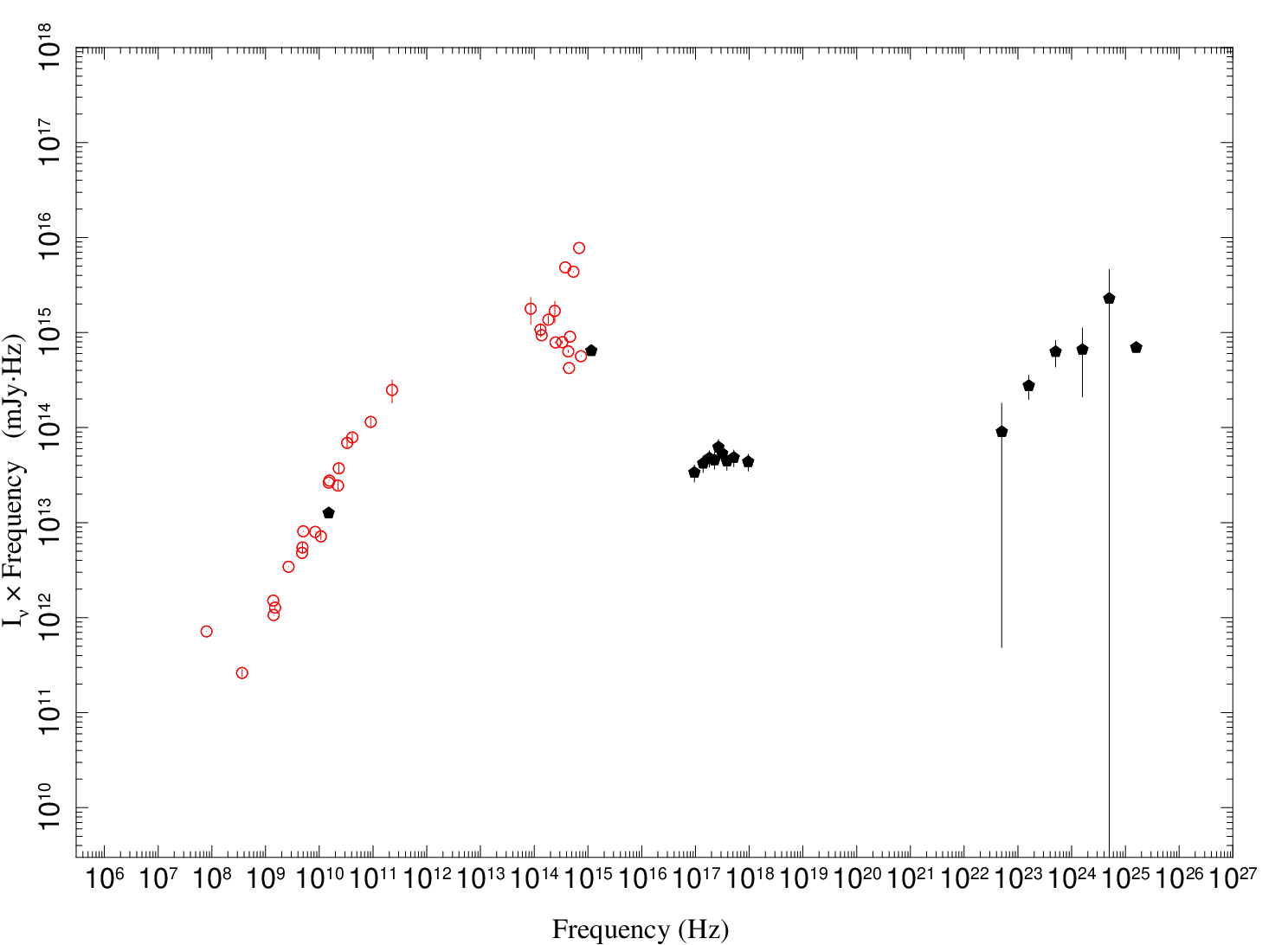}
 \caption{Broadband SED from radio to gamma-ray of PKS B0422+004. The data obtained before the \textit{Fermi} launch are red open circles (mostly taken from the NED), while the ones obtained later are black filled pentagons. The black data points are from: UMRAO, MOJAVE, \textit{Swift} UVOT/XRT, and \textit{Fermi} LAT. } 
 \label{fig:sed5}
\end{figure*}

\begin{figure*}
 \centering
 \includegraphics[angle=270, width=0.75\textwidth]{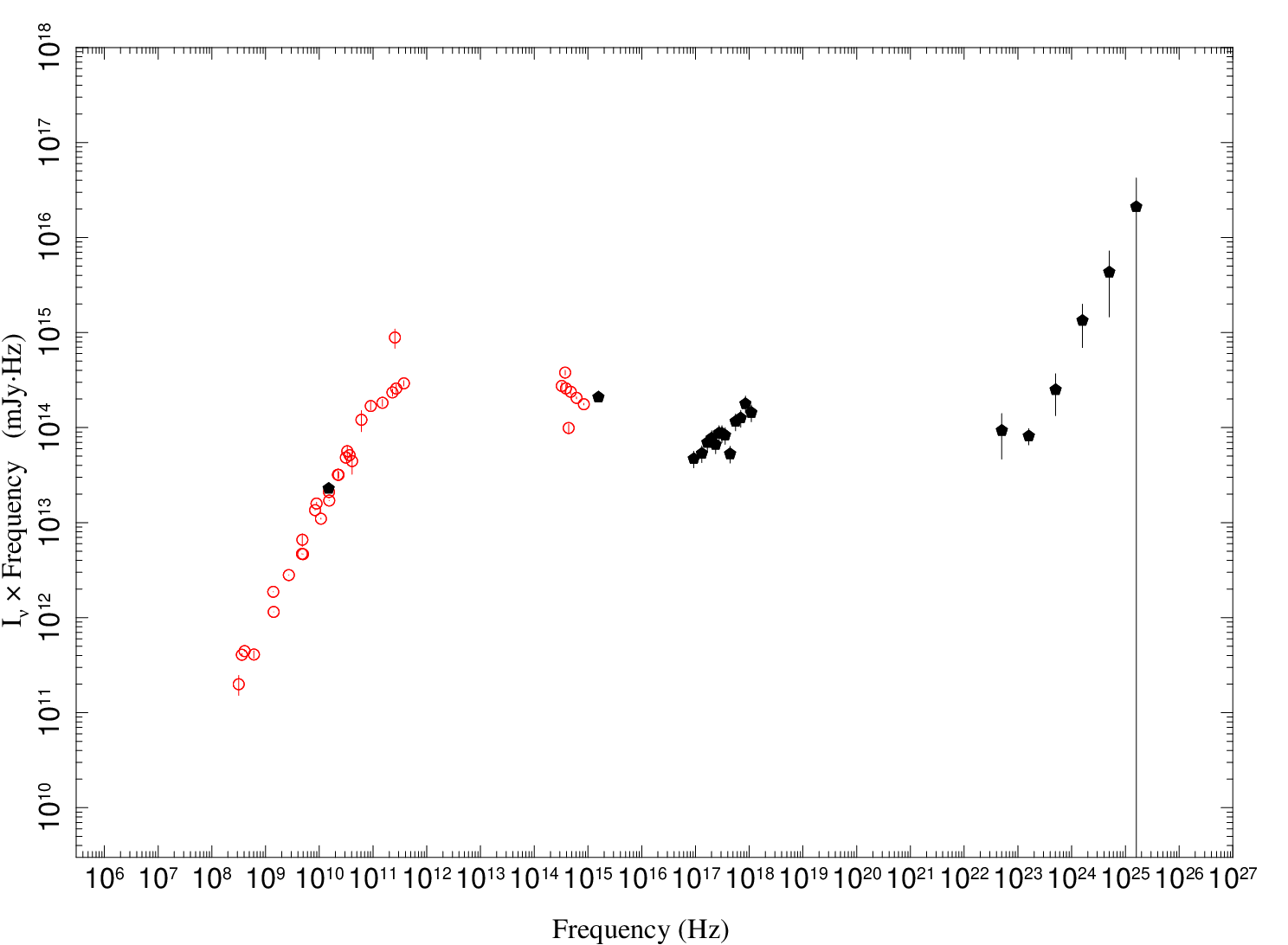}
 \caption{Broadband SED from radio to gamma-ray of PKS\,0823+033. The data obtained before the \textit{Fermi} launch are red open circles (mostly taken from the NED), while the ones obtained later are black filled pentagons. The black data points are from: UMRAO, MOJAVE, \textit{Swift} UVOT/XRT, and \textit{Fermi} LAT. }
 \label{fig:sed6}
\end{figure*}

\begin{figure*}
 \centering
 \includegraphics[angle=270, width=0.75\textwidth]{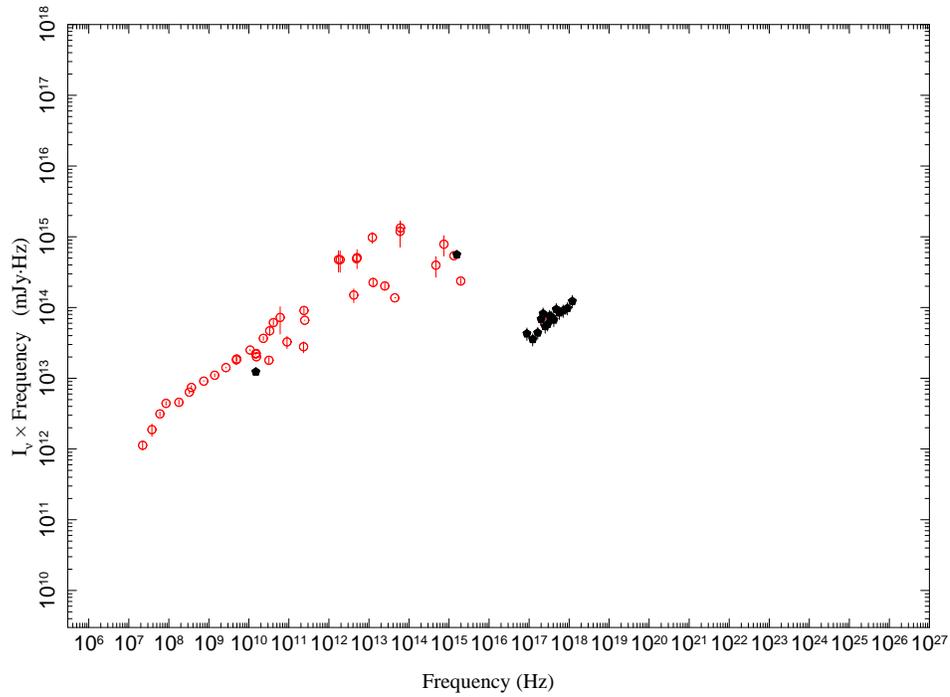}
 \caption{Broadband SED from radio to X-ray of 3C 309.1. The data obtained before the \textit{Fermi} launch are red open circles (mostly taken from the NED), while the ones obtained later are black filled pentagons. The LAT data are not included here because this source is not in the preliminary LAT first-year catalog. }
 \label{fig:sed1}
\end{figure*}

\begin{figure*}
 \centering
 \includegraphics[angle=270, width=0.75\textwidth]{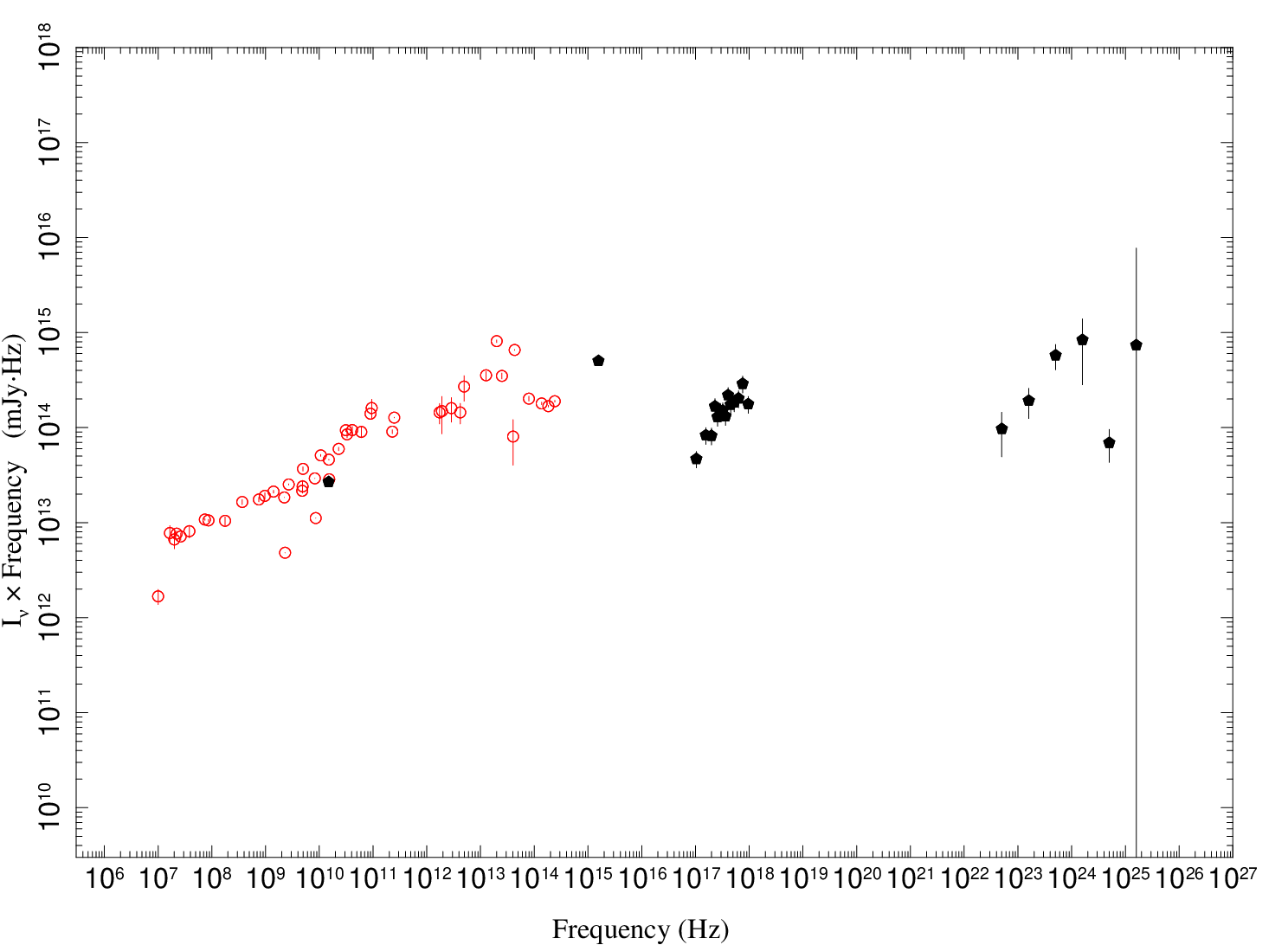}
 \caption{Broadband SED from radio to gamma-ray of 3C 380. The data obtained before the \textit{Fermi} launch are red open circles (mostly taken from the NED), while the ones obtained later are black filled pentagons. The black data points are from: UMRAO, MOJAVE, \textit{Swift} UVOT/XRT, and \textit{Fermi} LAT. }
 \label{fig:sed4}
\end{figure*}

Here we present preliminary broadband SED of the six sources from radio to gamma-ray energies (Figure \ref{fig:sed2} to \ref{fig:sed4}). The LAT spectral data presented here are obtained from the first 11-month dataset. We do not include the LAT spectrum of 3C 309.1 because it is not included in the preliminary \textit{Fermi} LAT year-one source list (Abdo et al., in preparation). 


There are two categories of SED models, with an interpretation of the high-energy part as being caused by Compton upscattering by leptons \citep{bottcher07} or by hadronic processes \citep{atoyan03}. The lower energy part (corresponding to the synchrotron radiation) is originated by leptons. For gamma-faint sources, we are interested in investigating the relation between the synchrotron peak and the synchrotron self-Compton (SSC) peak, and the extrapolation from the rising part before the SSC peak (X-ray) to the gamma-ray range. In this aspect, SED fitting is needed, and will be included in a near future. Here we discuss our preliminary SED results.

Most of the sources appear to have a comparable value of the synchrotron peak and the SSC peak. The spectra of NRAO 140, PKS\,0403$-$13, PKS\,B0422$+$004, and PKS\,0823+033 (Figure \ref{fig:sed2} to \ref{fig:sed6}) are flat to inverted in flux density, and the latter two are steeper. The slopes of the radio spectra are milder for 3C\,309.1 and 3C\,380 (Figure \ref{fig:sed1} and \ref{fig:sed4}). If we extrapolate the SED from X-ray to gamma-rays, we see that the SSC peak seems lower than the synchrotron peak in PKS\,B0422+004 and 3C\,309.1 (Figure \ref{fig:sed5} and \ref{fig:sed1}). For PKS\,0823$+$033, although the SSC peak appears slightly lower than the synchrotron peak by extrapolating from the X-ray spectrum, the gamma-ray flux increases at higher energies. This source is not in the TeV catalog \footnote{\texttt{http://tevcat.uchicago.edu/}}, and we need further analysis on the hardness of the gamma-ray spectrum in order to confirm this spectral rise in higher energies. For NRAO\,140 (Figure \ref{fig:sed2}), PKS\,0403$-$13 (Figure \ref{fig:sed3}), and 3C\,380 (Figure \ref{fig:sed4}), the SSC peak is comparable with the synchrotron peak. For gamma-bright blazars, the SSC peak in their SED tend to be a bit higher than the synchrotron peak (see \citep{abdo09}). In contrast to this, we see a trend that the SSC peak is comparable or lower than the synchrotron peak in the SED of these gamma-faint blazars, which are among the radio-brightest blazars in the sky. However, we need additional data to study the correlation between both peaks, as well as a detailed model of the physical process to reproduce the observed SED.

\bigskip

\section{Outlook}
We are compiling a broadband SED catalog for the 135 sources in the MOJAVE sample. Once collected, detailed analysis of the emission processes involved in the SED will be performed, and the deduced properties will be compared with the radio properties determined by the MOJAVE and by single-dish observing programs. Among the radio-bright sample, the gamma-faint sources are especially interesting for that they provide us a chance to probe the link between radio and gamma-rays, and to find out the key which affects the emission process. For further investigation, the broadband SED modelling will be performed in the future, in order to understand the nature of blazars. 


\bigskip 
\begin{acknowledgments}
We thank especially C. Ricci, M. B\"ock, L. Barrag\'an, J. Wilms, and C. M. Fromm for valuable discussions. This research was supported by the EU Framework 6 Marie Curie Early Stage Training program under contract number MEST/CT/2005/19669 “ESTRELA”. C. S. Chang is a member of the International Max Planck Research School for Astronomy and Astrophysics. This research includes data from observations with the 100-m telescope of the MPIfR (Max-Planck-Institut f\"ur Radioastronomie) at Effelsberg. This research has made use of data from the MOJAVE database that is maintained by the MOJAVE team \cite{lister09a}. This work is based on observations carried out with the IRAM Plateau de Bure Interferometer. IRAM is supported by INSU/CNRS (France), MPG (Germany) and IGN (Spain). The National Radio Astronomy Observatory is a facility of the National Science Foundation operated under cooperative agreement by Associated Universities, Inc. This research has made use of the NASA/IPAC Extragalactic Database (NED) which is operated by the Jet Propulsion Laboratory, California Institute of Technology, under contract with the National Aeronautics and Space Administration. This research has made use of data from the University of Michigan Radio Astronomy Observatory which has been supported by the University of Michigan and by a series of grants from the National Science Foundation, most recently AST-0607523.
\end{acknowledgments}

\bigskip 


\end{document}